# Production of Milky Way structure by the Magellanic Clouds[1]

Martin D. Weinberg[2]
Department of Physics & Astronomy
University of Massachusetts

## ABSTRACT

Previous attempts at disturbing the galactic disk by the Magellanic Clouds relied on direct tidal forcing. However, by allowing the halo to actively respond rather than remain a rigid contributor to the rotation curve, the Clouds may produce a wake in the halo which then distorts the disk. Recent work reported here suggests that the Magellanic Clouds use this mechanism to produce disk distortions sufficient to account for both the radial location, position angle and sign of the HI warp and observed anomalies in stellar kinematics towards the galactic anticenter and LSR motion.

*Subject headings:* galaxies: kinematics and dynamics, interactions, Milky Way galaxy, Magellanic Clouds

## 1. Introduction

All components of spiral galaxies are coupled by their mutual gravitational field. The fact that we only directly observe the luminous components focuses attention on the disk alone in attempts to understand their striking structure. For instance, "grand design" is often correlated with a companion or interloping galaxy and understood to be a result of differential or *tidal* acceleration of the disk. Similarly, researchers have tried to implicate our own fairly massive pair of companions, the Magellanic Clouds, in producing the observed warp in the Milky Way's disk. However the tidal force, which scales as the inverse distance cubed, is nearly an order of magnitude too small to do the job (e.g. Binney 1992).

---

The halo was postulated more than 20 years ago as an agent of dynamical stability for the disks of spiral galaxies (Ostriker & Peebles 1973) and data continue to imply the existence of this unseen mass. True to the original intent, most dynamical theories regard the halo as inert, only providing gravitational support for the luminous components. However the estimated mass of the dark halo is comparable to that of the disk inside of the Sun's galactocentric radius and dominates at larger radii. Therefore, any structure in the halo must surely affect the disk, and in this letter, I implicate the dark halo as a co-conspirator for exciting disk structure.

In short, the long-range gravitational response of the halo response can carry the external disturbance by the Magellanic Clouds to sufficiently small radii that the luminous disk can be affected (§2) and the disk response is predicted using perturbation theory (§3). Previous work has demonstrated that interactions between galaxies and between components of galaxies can be studied perturbatively with good results. Here, the theory is extended to include the combined effects of a disk embedded in a combined luminous and dark-matter halo. The implications and observational comparisons are discussed in §4.

## 2. Effect of the Clouds of the halo/spheroid component

Mass estimates for the Large Magellanic Cloud (hereafter LMC) range from $6 \times 10^9 \, M_\odot$ (Meatheringham et al. 1988) to $1.5 \times 10^{10} \, M_\odot$ (Schommer et al. 1992) which is from 10% to 30% of the stellar mass of the Milky Way's disk (following Binney & Tremaine 1987). Because the Clouds are distant, $R_g \approx 50 \, \mathrm{kpc}$, their trajectory traces the extent of the dark halo and considerable modeling and observational effort have been put to the task. Recently, the space velocity of the LMC has been redetermined from radial velocity and proper motions (Jones et al. 1994). Assuming a spherical halo, one may straightforwardly derive the following instantaneous orbit: $-76 \pm 13°$ inclination, $-82 \pm 10°$ longitude of ascending node, $-36 \pm 3°$ argument of perigalacticon[3].

The dark halo causes the Clouds orbit to decay by *dynamical friction*: our galaxy will eventual absorb the Clouds (Murai & Fujimoto 1980, Lin & Lynden-Bell 1982, Lin et al. 1995). The response of the dark halo to the interloping satellite can be thought of as a gravitational wake (e.g. Mulder 1983); since the wake trails and has mass, it exerts a backward pull on the satellite. This view of dynamical friction reproduces the standard results but can include the self-gravity (Weinberg 1989) and most important for our purposes here, gives us a way to estimate the distortion to the galaxy itself.

---

[3]The quoted errors are propagated by Monte Carlo analysis from the quoted standard errors in distance and proper motion but dominated by the latter.

Specifically, the response can be calculated by solving the collisionless Boltzmann equation (CBE) in the continuum limit. Since the satellite is small compared to the main galaxy, a perturbation expansion is a fair predictor of the true disturbance. The general approach has been tested and compared with n-body simulations in a variety of contexts with good agreement (e.g. Hernquist & Weinberg 1989, Weinberg 1993) and will be used here to consider both the effect of the satellite on the dark halo and luminous disk.

For an example, consider a dwarf companion in an extended halo[4]. Let the companion be a mass point on an eccentric orbit ($\epsilon = 0.4$) and currently at pericenter with coordinates $(X, Y) = (-50\,\mathrm{kpc}, 0\,\mathrm{kpc})$ similar to the LMC. Let the halo be a $W_0 = 7$ King model with $r_T = 100\,\mathrm{kpc}$. This choice gives a King core radius of roughly $3\,\mathrm{kpc}$ and a rotation curve that remains flat well-beyond the solar circle[5].

The $m = 1$ and $m = 2$ density components of the resulting wake are shown in Figures 1 and 2. The eccentric LMC orbit has two orbital frequencies whose harmonics determine resonant regions in halo phase space which couple strongly and non-locally. The wake peaks near the half-mass radius of the halo where the stellar orbital frequencies are changing quickly with radius. *Most importantly, the orbiting satellite can influence the inner galaxy via the global gravitational response of the halo which can carry the disturbance inward to the luminous disk!*

## 3. Combined effect of the Clouds and halo on the galactic disk

To estimate the amplitude of the disk response to the halo and LMC together, I extended the perturbative solution of the CBE to include a flat disk embedded in the dark halo. The disk disturbance is confined to the plane. Both the disk and halo react gravitationally to each other, themselves, and the external satellite. The initial disk has the exponential density profile typical of spirals; the distribution function is constructed self-consistently by a quadratic programming technique (Dejonghe 1989) penalized to prefer a cold tangential distribution with an appropriate velocity dispersion. The gravitational field of each component is represented by a truncated series of orthogonal functions (e.g. Clutton-Brock 1972, 1973, Hernquist & Ostriker 1992). With this discretization, the response is a solution to a matrix equation; the individual galaxian components are gravitationally coupled by matrices which transform one set of orthogonal functions to all others. Since an orthogonal polar disk component, proportional to $\exp(im\phi)$, couples to all spherical components $Y_{lm}$ with $l \geq m$, all desired values $l$ for a single $m$ must be considered together. Once the perturbed distribution function is known, all observable quantities such as density and line-of-sight velocity are immediate consequences.

---

[4]I will refer to total non-disk distribution of mass as *halo*; the halo, then, contains the traditional bulge, luminous spheroid and extended dark matter components.

[5]Below, we will add an exponential disk with mass comparable to the halo inside the solar radius which gives a fair fit to the observed rotation curve.

Here we consider the low-order $m = 1, 2$ components. Higher-order harmonics do contribute but their effect is likely to be over estimated due to the assumption of a slowly and smoothly evolving LMC orbit. A detailed discussion will be presented in a later paper. The $m = 1$ disturbance shifts the center of mass. Since the LMC is in the outer halo, the dominant effect on the halo will be a center-of-mass or barycentric shift (Weinberg 1989). This is not true for the disk for two reasons: 1) LMC orbital plane is highly inclined and the projection of the $l = m = 2$ component onto the disk has an $m = 1$ contribution; and 2) the disk is responds more strongly to the halo and its distortions than the LMC tide.

Figures 3 and 4 depict the distortions of the disk by the halo and the LMC with the orbit plane tilted and rotated as stated in §2. In both the $m = 1$ and $m = 2$ cases, the dominant response is near or outside the solar circle and has significant amplitude: several percent at the solar circle reaching 30% at $R_g \approx 15\,\mathrm{kpc}$. The LMC has a retrograde orbit relative to disk rotation, and the response has a negative pattern speed relative to the disk rotation. Therefore, the slow retrograde patterns in Figures 3–4 can not be directly responsible for classic bar morphology but could be the triggering source.

## 4. Implications and summary

### 4.1. Comparison with inferred Milky Way asymmetries

Stellar spectroscopy of K–giants and carbon stars (Lewis & Freeman 1989, Metzger & Schechter 1994) indicate a net streaming motion relative to the LSR toward the galactic anticenter of approximately $10\,\mathrm{km\,s^{-1}}$. An axisymmetric galaxy would have no streaming. Most explanations for the implied non-axisymmetries are based on weak oval or quadrupole distortions. It is clear from Figures 3 and 4 that the disk response to the LMC engendered halo wake may bear on this interpretation. Figure 5 shows the predicted kinematic asymmetries outside the solar circle. The $m = 1$ component is nearly constant and ingoing at $10\,\mathrm{km\,s^{-1}}$. Superimposed on this, the $m = 2$ component causes a recession of the outer galaxy relative to the Sun's position. The combined line-of-sight velocity as seen from the galactic center is shown as a solid line with the solar circle indicated. Relative to the LSR, the outer galaxy recedes at the observed $10\,\mathrm{km\,s^{-1}}$ and agrees with the stellar kinematics within the measurement errors. The model also implies a net $12\,\mathrm{km\,s^{-1}}$ inward velocity for the LSR. However, the observed kinematic center must be compared with the apparent kinematic center of the perturbed model not the original unperturbed center. This recentering may be estimated by removing the $11\,\mathrm{km\,s^{-1}}$ velocity offset which leaves, roughly, the $m = 2$ signature alone.

Blitz & Spergel (1991) use a quadrupole distortion deduce an *outgoing* LSR based on HI kinematics. The time-dependent LMC wake might easily cause a transient gas response which differs from the stellar response but has similar amplitude. More generally, these estimates suggest observable velocity signatures near and outside the solar circle caused by the Magellanic Clouds, but more detailed modeling will be required for precise predictions.

### 4.2. Implications for the HI warp

Figure 2 shows that a strong halo overdensity occurs at $R_g \approx 15\,\mathrm{kpc}$ which coincides with the peak of the observed warp. For reference, its peak distortion is towards $l = 90°$ and positive $b$ at $R_g \approx 16\,\mathrm{kpc}$ and towards $l = 270°$ and negative $b$ at $R_g \approx 15\,\mathrm{kpc}$. We focus on the $m = 2$ component because a large fraction of the $m = 1$ distortion will be absorbed in the center-of-mass shift. After appropriate rotation of the orbital plane, the gravitational potential corresponding to the $m = 2$ wake has two minima at $X = \mp 1\,\mathrm{kpc}, Y = \pm 15\,\mathrm{kpc}$ with the peak at positive (negative) $Y$ above (below) the galactic plane. The disk model described here is planar and can not predict the amplitude of the warp. However, Figure 4 shows that the disk distortion is $\gtrsim 30\%$ at $X = \mp 1\,\mathrm{kpc}, Y = \pm 15\,\mathrm{kpc}$, suggesting that significant vertical distortion is likely. The orientation of the LMC orbital plane is such that the Northern (Southern) distortions will be above (below) the plane as observed. As emphasized by Binney (1992), the halo will play a dominant rôle in warp dynamics. Although the Clouds can not drive the warp directly, this work suggests that an *active* halo distorted by the Clouds *can* drive the warp.

### 4.3. Future work

Clearly much remains to be done before the full implications of this mechanism is understood. The following three broad topics are ripe for rapid progress:

- *Effect of the three-dimensional structure of disk.* This is a straightforward application for n-body simulation but will require very large numbers of particles. Alternatively, one could use a hybrid approach with an analytic model for the halo distortion together with an n-body disk.

- *Gas response and comparison with observed HI.* It is likely that the time-dependent effects are crucial to understanding the stellar and gas kinematics together. Both periodic orbit analyses and direct n-body simulations with gas could be performed and compared.

- *Non-linear development.* The results presented here describe a forced response only. The slow retrograde distortions might be amplified by the standard mechanisms to drive spiral structure and perhaps inner bars.

This work was supported in part by NASA grant NAG 5-2873.

– 7 –

# FIGURE CAPTIONS

Fig. 1.— The density wake in the orbital plane for $m = 1$ summed over harmonic components $l = 1, \ldots, 6$. Linearly spaced contours of overdensity (shaded). Underdensity is identical but rotated by 180°. The satellite is on an eccentric orbit ($\epsilon = 0.4$) in the $X$–$Y$ plane and is at pericenter at $X = -50\,\mathrm{kpc}, Y = 0\,\mathrm{kpc}$ and is moving the $-\hat{Y}$ direction. The dashed circle denotes the solar radius.

Fig. 2.— As in Fig. 1 but for $m = 2$. Underdensity is identical but rotated by 90°. The two-armed pattern trails the satellite.

Fig. 3.— Linear contours of positive overdensity for the $m = 1$ disk response (shaded). The dashed circle indicates the solar radius; the Sun has coordinates $(-8.5, 0)$. Logarithmic contours of relative underdensity (dotted). Levels: 1, 3, 10, 30, 100%. The 1% contour intersects the solar circle.

Fig. 4.— As in Fig. 3 but for $m = 2$.

Fig. 5.— Line-of-sight velocity for the $m = 1$ (dotted), $m = 2$ (dashed), and combined (solid) responses viewed from the galactic center. Values are scaled to the Milky Way: scale length $a = 3.5\,\mathrm{kpc}$ and rotation speed $V_o = 220\,\mathrm{km\,s^{-1}}$. The IAU solar position of $R_g = 8.5\,\mathrm{kpc}$ is indicate by the dotted vertical line.